\documentclass[baaa]{baaa}

\usepackage[pdftex]{hyperref}
\usepackage{subfigure}
\usepackage{natbib}
\usepackage{helvet,soul}
\usepackage[font=small]{caption}

\contriblanguage{1}

\contribtype{1}

\thematicarea{1}

\title{Photometric and Morphological Analysis of Fornax Galaxies through S-PLUS}

\titlerunning{Analysis of Fornax Galaxies with S-PLUS}

\author{
A.V. Smith Castelli\inst{1,2},
A.R. Lopes\inst{3},
A. Cortesi\inst{4,5},
P. Dimauro\inst{3},
R. Cid Fernandes\inst{6},
G. Lucatelli\inst{7,8},\\
C. Mendes de Oliveira\inst{8},
F. Almeida-Fernandes\inst{8,9},
J. T. S. C. Batista\inst{6},
D. Brambila\inst{5},
S. Dhiwar\inst{10},\\
P. Lopes\inst{5}
\&
K. Saha\inst{10}
}

\authorrunning{Smith Castelli et al.}

\contact{asmith@fcaglp.unlp.edu.ar}

\institute{
Insituto de Astrof\'isica de La Plata, CONICET--UNLP, Argentina
\and
Facultad de Ciencias Astron\'omicas y Geof\'isicas, UNLP, Argentina
\and
Observatorio Nacional, Brasil
\and
Centro Brasileiro de Pesquisas F\'isicas, Brasil
\and
Observatorio  do Valongo, UFRJ, Brasil
\and
Departamento de F\'isica, UFSC, Brasil
\and
Jodrell Bank Centre for Astrophysics, UM, Reino Unido
\and
Instituto de Astronomia, Geof\'isica e Ci\^encias Atmosf\'ericas, USP, Brasil
\and
Community Science and Data Center/NSF’s NOIRLab, Estados Unidos
\and
The Inter-University Centre for Astronomy and Astrophysics, India
}

\resumen{El an\'alisis fotom\'etrico y morfol\'ogico de las poblaciones de galaxias en c\'umulos provee informaci\'on valiosa acerca del estado evolutivo del c\'umulo al que pertenecen. Asimismo, permite comprender la influencia del medioambiente en las propiedades observadas de las galaxias y, como consecuencia, sus caminos evolutivos. En esta contribuci\'on presentamos los primeros pasos en el an\'alisis fotom\'etrico y morfol\'ogico de las galaxias del c\'umulo de Fornax utilizando datos de S-PLUS. Esperamos que el novedoso conjunto de filtros y la cobertura de campo amplio de S-PLUS nos permita obtener nueva informaci\'on acerca de Fornax y de su poblaci\'on de galaxias. }

\abstract{The photometric and morphological analysis of galaxies in clusters provides invaluable information regarding the evolutionary stage of the cluster itself. In addition, it helps to understand how the environment affects the properties of the galaxies and, as a consequence, their evolutionary path. 
In this contribution we present the first steps on the photometric and morphological analysis of galaxies in the Fornax cluster using S-PLUS data. We expect that the S-PLUS novel filter set and wide field coverage allow us to obtain new information about Fornax and its galaxy population.
}


\begin{document}

\maketitle

\section{Introduction}
\label{intro}
The Southern Photometric Local Universe Survey (S-PLUS) aims at mapping $\sim$9300 deg$^2$ of the southern sky in the 5 broad-bands of the Sloan Digital Sky Survey (SDSS) and 7 narrow-bands tracing specific spectral features such as [OII], $\rm{H_\delta}$ and $\rm{H_\alpha}$ among others. One of the goals of S-PLUS is to deliver regular data releases. DR1, DR2 are currently available\footnote{https://splus.cloud/documentation/datareleases} and DR3 was recently announced. All the information related with the survey can be found in \citet{MendesOliveira2019}.

Fornax is the second nearby rich galaxy cluster after Virgo and it is observable from the South. Though highly studied, it has never been analyzed simultaneously in 12 photometric optical bands. One of the projects within the S-PLUS collaboration is the S-PLUS Fornax Project (S+FP)  which aims at analyzing the galaxy and globular cluster population of Fornax using S-PLUS data \citep{SmithCastelli2021}. It is expected that such a project  provides new and valuable information on the cluster. In this contribution we briefly present some of the topics that will be developed in the framework of the S+FP.

%The Southern Photometric Local Universe Survey (S-PLUS) is an international collaboration that started in 2016, aiming at imaging $\sim$9300 deg$^2$ of the southern sky in 12 photometric bands. All the information related with the survey can be found in \citet{MendesOliveira2019}. One of the projects within the collaboration is related with the analysis of the Fornax cluster. Though highly studied, Fornax has never been analyzed simultaneously in 12 photometric bands. We expect that such an analysis will provide new and valuable information on the cluster. The S-PLUS Fornax project (S+FP) was already introduced by \citet{SmithCastelli2021}

%\section{Section 1}

%\subsection{Subsection 1.1}

%\begin{figure}[!t]
%\centering
%\includegraphics[width=\columnwidth]{green_dots.png}
%\caption{Green dots in an spiral galaxy}
%\label{green_dots}
%\end{figure}

\section{Emission lines in different galaxy types}
\label{SF_EL}

\begin{figure}[!b]
    \centering
    \includegraphics[width=0.96\columnwidth]{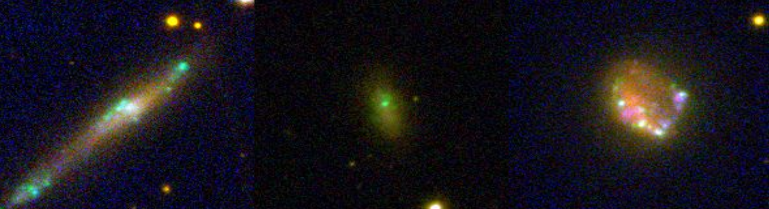}
    \caption{From left to right, 200 $\times$ 200 pixel$^2$ images in four S-PLUS photometric bands of ESO\,357-7, FCC\,35 and LEDA\,4080439. The green color display the contribution of $J0660$ (H$\alpha$), while the blue color combines the $u$ and $J0378$ ([OII]) bands. The red color comes from the $r$ filter. The blue-green regions correspond to star-forming regions. }
    \label{fig:halpha_oii_examples}
\end{figure}

At the distance of the Fornax cluster, [OII] and H$\alpha$ emission lines fall within the S-PLUS narrow-band filters $J0378$ and $J0660$, respectively. The presence of these lines are identified by an excess in colors $(u-J0378)$ and $(r-J0660)$, respectively. Figure\, \ref{fig:halpha_oii_examples} presents three examples of Fornax galaxies with emission. 

\begin{figure}
    \centering
    \includegraphics[width=0.96\columnwidth]{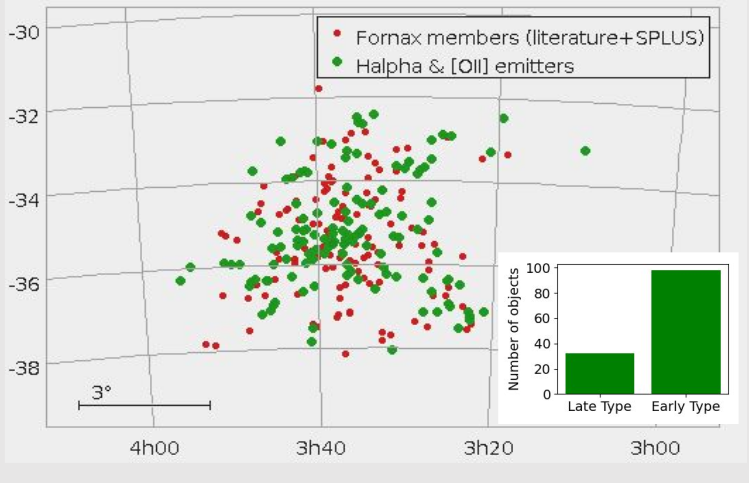}
    \caption{Spatial distribution of 258 Fornax members from the literature observed with S-PLUS. In green, we depict the 130 [OII] and H$\alpha$ emitters. The number of early and late-type galaxies with  emission is shown in the inner bar plot.}
    \label{fig:oii_halpha}
\end{figure}

Based on a sample of 258 objects, which are a cross-match between literature members of Fornax and S-PLUS observations, we perform a first selection of [OII] and H$\alpha$ emitters. We find 130 sources with excess in $(u-J0378)$ and/or $(r-J0660)$ colors. The spatial distribution of the 258 Fornax galaxies with confident S-PLUS photometry depicting in green the emitters, is shown in Figure\,\ref{fig:oii_halpha}. The bar plot in the bottom left of that figure indicates that most emitters are early-type galaxies. That is an interesting result that will be further developed in future papers. 

However, we still need to refine the selection of the emitters, as many of those objects display a high red continnum that can strongly influence the detection of the emission. In addition, we realized that S-PLUS total magnitudes might not be efficient to properly hint the presence of emission in galaxies at the distance of Fornax. A deeper analysis considering the size and morphology of the galaxy is ongoing.

Once the emitters list is concluded, we will derive the emission line fluxes following the {\it Three Filter Method} \citep[e.g.][]{Pascual2007,VilellaRojo2015}, and then estimate the star formation rates \citep[e.g.][]{Kennicutt1998}. Our goal is to establish a connection between the emitters and the different morphological types, so we can trace the star formation activity in Fornax galaxies and find additional information about the cluster evolution.  

\section{Spectral fitting and non-parametric star formation histories of Fornax galaxies}
\label{SED_SFH}

\begin{figure}[!t]
\centering
\includegraphics[width=\columnwidth]{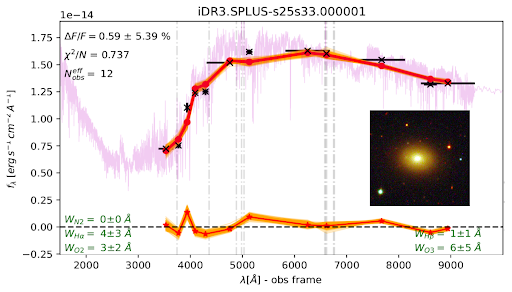}
\includegraphics[width=\columnwidth]{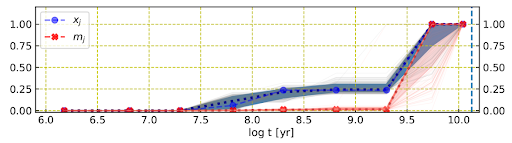}
\caption{{\it Top panel:} Example of spectral fitting with AlStar. The fit was performed on the S-pectrum of FCC\,43, a spectroscopically confirmed member of the Fornax cluster (small inset).  {\it Bottom panel:} The non-parametric SFH of FCC\,43 is shown. See details in the text. }
\label{SED}
\end{figure}

One of the topics to be developed within the S+FP is that of spectral fitting and the analysis of the star formation histories of the galaxies in the cluster. The top panel of Figure\,\ref{SED} shows the 12-band S-PLUS photometric spectrum (hereafter, S-pectrum) of FCC\,43, a spectroscopically confirmed member of the Fornax cluster, (black points) and the AlStar fit (red lines). Orange lines show fits to 100 perturbed versions of the data. The pink line in the background shows the best fit model spectrum in high resolution. The non-parametric star formation history (SFH) is modeled as a linear combination of stellar populations of 9 ages ($0 < t < 14$ Gyr) and 5 metalicities (from 1/3 to 3 solar). Emission lines are accounted for in the fits as well.

In the bottom panel of Figure\,\ref{SED}, the non-parametric SFH of FCC\,43 is shown as cumulative flux (blue) and mass (red) curves, ranging from 0 at $t=0$  to 1 at $t=14$ Gyr ago. The blue and red bands represent ranges of possible solutions, as obtained from  Monte Carlo runs. The dashed lines indicate the best fit to the observed data, while the dotted ones trace the mean solution of the simulated spectra. Most of the stellar mass in FCC\,43 reside in old populations. Stars younger than 1 Gyr also seem to be present, with a $\sim 25\%$ contribution in light but very little mass. Emission lines are essentially absent in this galaxy, as seen by the equivalent widths listed at the bottom of the plot \citep{Batista2021}.

%-- see J. Batista MSc thesis (at https://tede.ufsc.br/teses/PFSC0405-D.pdf) for more details on the code.

%{\bf\large You may want to cite 
%Batista, J. T. S. C., MSc thesis, https://tede.ufsc.br/teses/PFSC0405-D.pdf}

\section{Effect of the environment on galaxy evolution }
What are the main mechanisms that shape or have shaped galaxies to make them appear with the great variety of morphologies that we observe today, is one of the main open questions in astrophysics. It has been widely discussed whether and how the environment drives such evolution. Nowadays, there is general consensus on its importance. Indeed, structural properties as well as star formation activity of galaxies that reside in high density environments, as clusters or groups, are found to differ from the ones of the field counterpart (\citealp{Dressler1980,Balogh2004}). The Fornax cluster is one of the largest structure in the local Universe. Moreover, it is composed by a set of galaxies that present a large variety of proprieties, a characteristics that makes Fornax one of the best mediums to investigate the role of the environment on the evolution of galaxies.

\begin{figure}
\centering
\includegraphics[width=0.96\columnwidth]{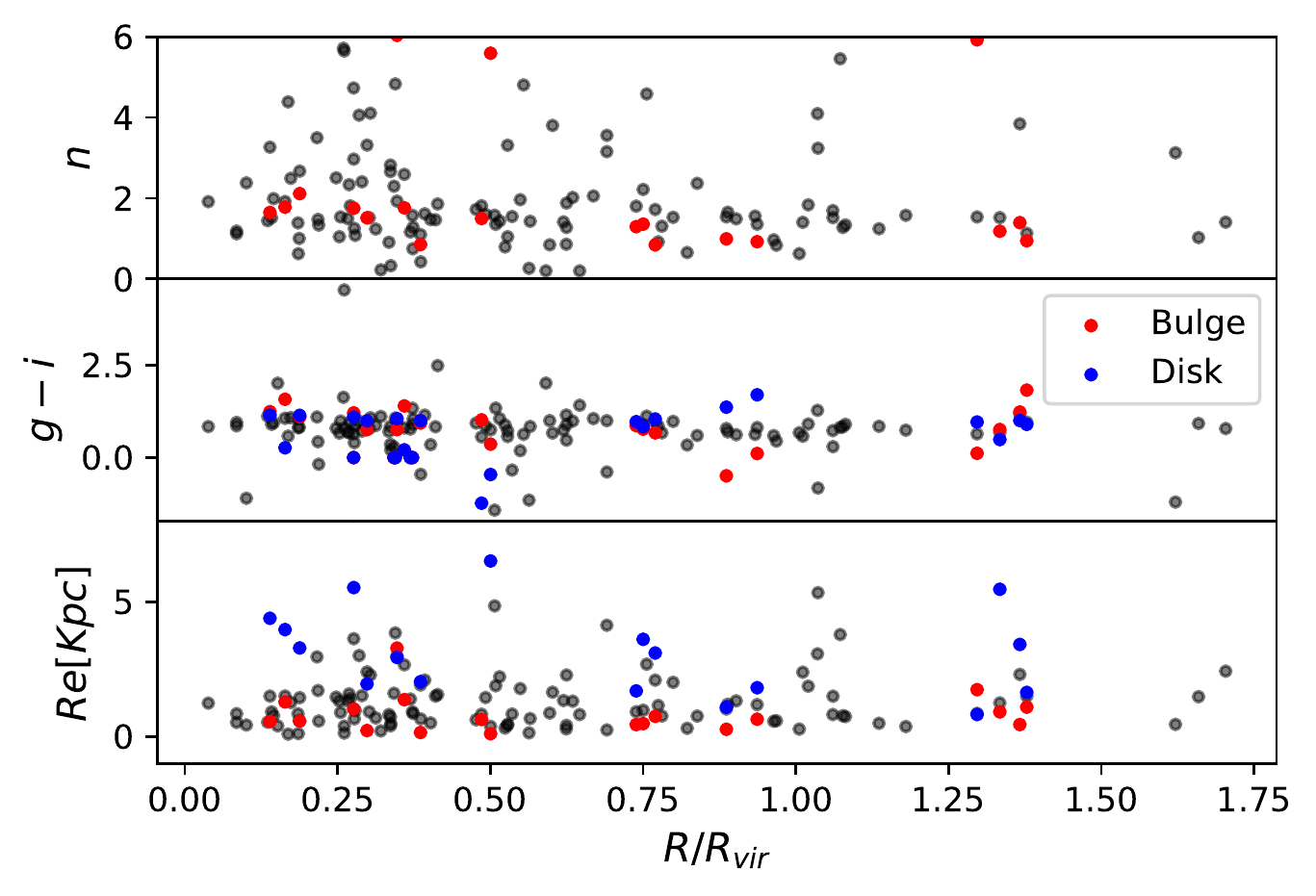}
\caption{Radial distribution of structural and stellar properties of galaxy members. {\it Top panel:} Radial distribution of the S\'ersic Index. {\it Mid and Bottom panels:} Radial distribution of, respectively,  half-light sizes and g-i colors of galaxies members but also of their internal components (bulges and discs). }
\label{Morph}
\end{figure}

Figure\,\ref{Morph} shows the radial distribution of structural and stellar properties of galaxy members, with the aim of finding signatures of the environment. Galaxy models were recovered running Galapagos-2 \citep{Galapagos-2} over the entire set of filters of S-PLUS. If the over-density, related to the center of the cluster, is playing a major role affecting the evolution galaxies within it, a gradient is expected in the radial distribution of both morphology and stellar properties. Results from Figure\,\ref{Morph} underline a weak correlation between galaxy properties and the position of the galaxy in the cluster, suggesting that the environment is not the major driver in their evolution.

\section{Blue Elliptical galaxies}
\begin{figure}
\centering
\includegraphics[width=0.96\columnwidth]{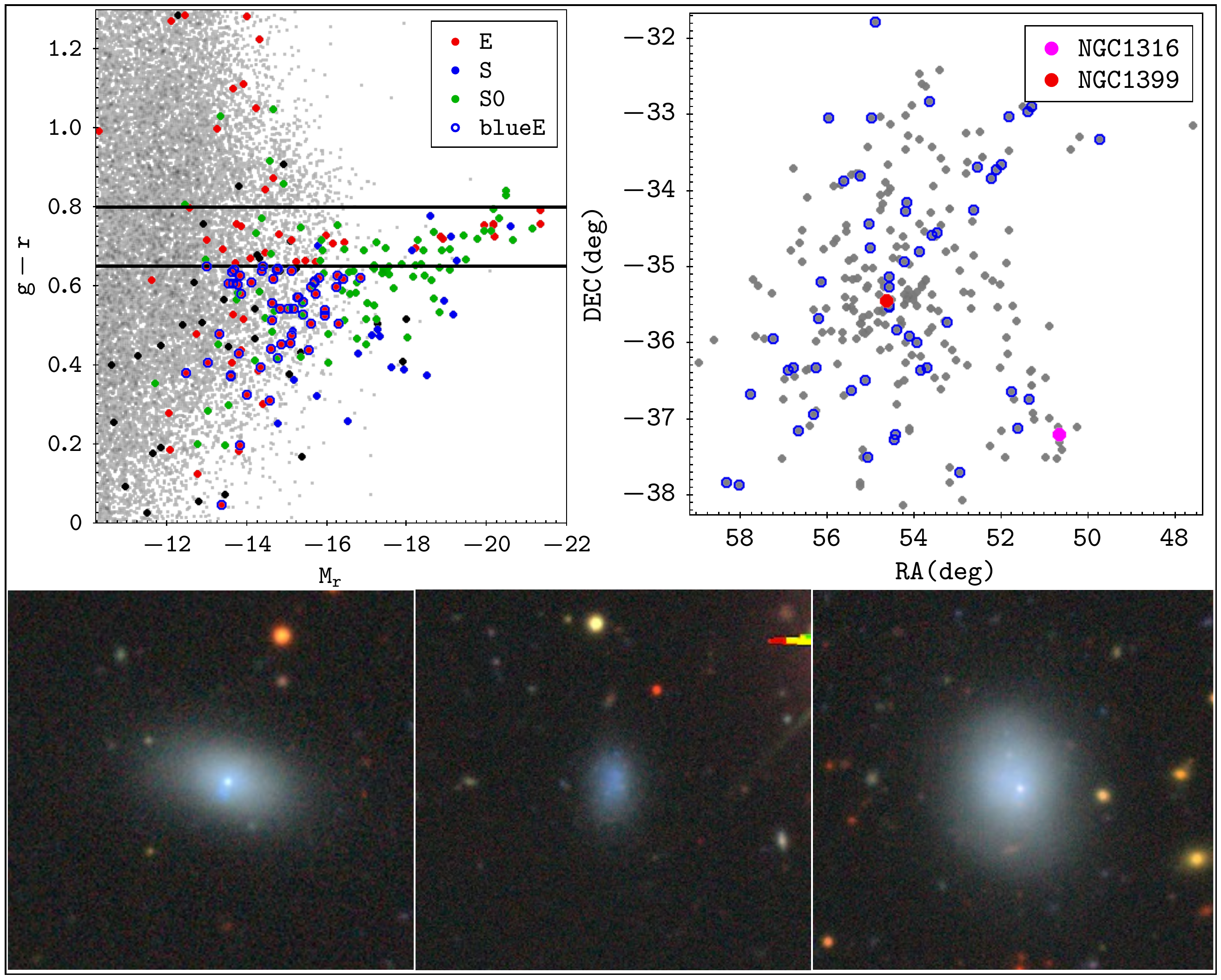}
\caption{{\it Top left panel:} colour-magnitude diagram of galaxies in this study, colour coded according to their morphology. The grey dots show S-PLUS sources with high probability of being galaxies. {\it Top right panel:} spatial distribution of the blue Es as well as the location of NGC\,1316 and NGC\,1399, the two dominant galaxies of Fornax. {\it Bottom panels:} Some examples of blue Es in Fornax. }
\label{blueE}
\end{figure}

Elliptical (E) galaxies present a smooth ellipsoidal morphology, reflecting the random and often very elongated orbits of their constituent stars. Generally, E galaxies live in the densest regions of the Universe \citep{Dressler1980}, i.e. in the centre of clusters and groups, and are characterised by an old stellar population, rendering their integrated colour red. 
Yet, blue E galaxies have been discovered (Suraj Dhivar et al. in prep.) and their formation is still unknown. At the same time, they are  a precious piece to compose the puzzle of galaxy formation and evolution. In fact, they might be the result of recent star formation, induced either from the environment the galaxy live in, or accretion. Of particular interest is studying their location in galaxy clusters, as the Fornax cluster. In this pilot study, we used a colour magnitude diagram built from S-PLUS photometry (see top left panel of Figure\,\ref{blueE}) and a colour-colour diagram (g-r vs r-i), to identify blue E galaxies in Fornax (note that the morphology is from the literature). Then, we visually inspected all the selected objects creating gri colour images from {\it The DESI Legacy Imaging Surveys}\footnote{https://www.legacysurvey.org/}, which has a higher resolution than S-PLUS (see the bottom panel in Figure\,\ref{blueE} for three examples). In the top right panel, we show the position of the galaxies within Fornax: blue E galaxies are depicted as dark blue dots and the location of the two brightest galaxies of the cluster (NGC\,1316 and NGC\,1399) is marked with a magenta and red dot, respectively. It is possible to see that some blue Es are located in the region between the two main Fornax sub-structures dominated by NGC\,1316 and NGC\,1399. That might suggest that, in those cases, the star formation could originate in the merger of the two sub-structures. 

%\section{Future work}
%In the context of the S+FP, we expect to extend all these analysis and publish the results in the near future. 

\begin{acknowledgement}
We thank the anonymous referee for her/his report that helped to improve the content of this contribution. S-PLUS is an international collaboration founded by Universidade de Sao Paulo, Observat\'orio Nacional, Universidade Federal de Sergipe, Universidad de La Serena and Universidade Federal de Santa Catarina.
This work was funded with grants from FAPESP, CONICET, Agencia I+D+i and Universidad Nacional de La Plata.

\end{acknowledgement}

%%%%%%%%%%%%%%%%%%%%%%%%%%%%%%%%%%%%%%%%%%%%%%%%%%%%%%%%%%%%%%%%%%%%%%%%%%%%%%
%  ******************* Bibliografía / Bibliography ************************  %
%                                                                            %
%  -Ver en la sección 3 "Bibliografía" para mas información.                 %
%  -Debe usarse BIBTEX.                                                      %
%  -NO MODIFIQUE las líneas de la bibliografía, salvo el nombre del archivo  %
%   BIBTEX con la lista de citas (sin la extensión .BIB).                    %
%                                                                            %
%  -BIBTEX must be used.                                                     %
%  -Please DO NOT modify the following lines, except the name of the BIBTEX  %
%  file (whithout the .BIB extension).                                       %
%%%%%%%%%%%%%%%%%%%%%%%%%%%%%%%%%%%%%%%%%%%%%%%%%%%%%%%%%%%%%%%%%%%%%%%%%%%%%% 

\bibliographystyle{baaa}
\small
\bibliography{Morphology_Fornax}
 
\end{document}